# Three-dimensional phonon population anisotropy in silicon nanomembranes

Kyle M. McElhinny[1], Gokul Gopalakrishnan[1†], Martin V. Holt[2], David A. Czaplewski[2], Paul G. Evans[1*]

[1] Materials Science and Engineering, University of Wisconsin-Madison, Madison, WI 53706, USA

[2] Center for Nanoscale Materials, Argonne National Laboratory, Argonne, IL 60439, USA

* Corresponding author. E-mail: pgevans@wisc.edu

†Present address: Engineering Physics, University of Wisconsin-Platteville, Platteville, WI 53818, USA

**Abstract**: Nanoscale single-crystals possess modified phonon dispersions due to the truncation of the crystal. The introduction of surfaces alters the population of phonons relative to the bulk and introduces anisotropy arising from the breaking of translational symmetry. Such modifications exist throughout the Brillouin zone, even in structures with dimensions of several nanometers, posing a challenge to the characterization of vibrational properties and leading to uncertainty in predicting the thermal, optical, and electronic properties of nanomaterials. Synchrotron x-ray thermal diffuse scattering studies find that freestanding Si nanomembranes with thicknesses as large as 21 nm exhibit a higher scattering intensity per unit thickness than bulk silicon. In addition, the anisotropy arising from the finite thickness of these membranes produces particularly intense scattering along reciprocal-space directions normal to the membrane surface compared to corresponding in-plane directions. These results reveal the dimensions at which calculations of materials properties and device characteristics based on bulk phonon dispersions require consideration of the nanoscale size of the crystal.

## I. INTRODUCTION

Recent advances in materials fabrication provide an increased scope for the control of vibrational properties through the engineering of nanoscale structures and periodic boundaries.[1–3] The control of phonon propagation in solids is advancing towards the level already achieved with electrons and photons.[3] Phonon scattering rates and the dispersions differ significantly between bulk materials and nanoscale materials,[4–6] and structures with periodicities ranging from centimeters to nanometers can be created to manipulate phonons at frequencies relevant to the propagation of sound, ultrasound, GHz acoustics, and heat.[3,7,8]

The vibrational parameters of many bulk materials have been well resolved for several decades, thanks largely to inelastic spectroscopy techniques in neutron and x-ray scattering.[9,10] Nanomaterials, however, pose a significant and ongoing challenge due to their small volumes and, as a result, the characterization of phonons in nanomaterials has been largely accomplished using optical techniques. Nanostructures formed from Si and Si-alloy semiconductors illustrate the limitations of the optical approach and have widespread thermal, electronic, and optical applications. The momenta of the THz-frequency phonons relevant to thermal transport in Si are far larger than can be probed optically, reaching corresponding wavevectors on the order of 1 Å$^{-1}$.[11–16]

Large-wavevector acoustic phonons are particularly important in electronic processes in Si because electron-phonon coupling involves phonons with wavevectors connecting widely separated conduction band minima (at the six $2\pi/a_{Si}$ [0 0 0.85] reciprocal lattice points, with lattice parameter $a_{Si}$ = 5.43 Å), with wavelengths on the order of a nanometer.[17] However, the length scale at which the energies and density of states at these wavevectors deviate from the bulk values is presently unknown. In addition to the drastic modification of the phonon dispersion expected in structures with dimensions matching the relevant phonon wavelengths (i.e., around a nanometer), it is particularly important to know at what length scales to expect deviations from bulk dispersions that are smaller but still important to the proper modeling of electrical and thermal properties. This is especially relevant to understanding not just electron-phonon scattering, but also heat dissipation and thermoelectric performance in devices that typically have dimensions of a few tens of nanometers.

The magnitude of the characterization challenge can be made conceptually clear by dividing the Brillouin zone characterizing the momentum space of Si into two parts: a small-wavevector region near the origin of reciprocal space, and a surrounding region spanning a far larger volume of reciprocal space characterized by wavevectors with magnitudes near the inverse of the lattice parameter. The small-wavevector regime of the phonon dispersion in nanomaterials with wavevectors less than $10^{-3}$ Å$^{-1}$, is strongly modified in comparison with the bulk, and has been extensively probed by optical techniques such as Raman and Brillouin scattering.[18–25] The region probed by optical scattering with visible or UV photons, however, spans just a small fraction, approximately $(10^{-3}/1$ Å$^{-1})^3 \approx 10^{-9}$, less than a billionth of the relevant volume of reciprocal space. The vast majority of phonon modes in nanomaterials thus remain uninvestigated by optical techniques. Molecular dynamics studies of nanoscale membranes predict that modes with low frequencies extend across the Brillouin zone. Predictions for a nanoscale Si sheet, particularly important in both thermal and electronic transport, reveal acoustic modes with frequencies reduced by approximately 50% at the zone edge.[5] In this work, we extend the range of phonon wavevectors in nanoscale systems that can be studied by scattering techniques by probing the population of phonons in a three-dimensional momentum-space volume over an extremely wide range of wavevectors, extending from near the zone center to the very edge of the Brillouin zone. These synchrotron x-ray thermal diffuse scattering (TDS) measurements probe phonon populations across the entire three-dimensional Brillouin zone of a silicon nanomembrane with a thickness of 21 nm, and expose a significant anisotropy in the phonon population.

The vibrational properties of nanoscale crystals involve effects arising both from the atomic-scale crystal structure and from the overall shape of the crystal. The frequencies and displacements associated with the vibrational modes of nanomaterials have a complex spatial dependence because spatial confinement lifts the requirement that modes be periodic. Crystallographic directions that are equivalent in the bulk (e.g. [001], [100], and [010]) can exhibit anisotropy in low-dimensional nanostructures. The resulting vibrational modes of nanomaterials involve atomic displacements for which the bulk description in terms of longitudinal and transverse polarizations is not strictly correct.[26–28] The in-plane and out-of-plane directions in a nanoscale sheet, for example, have distinct features in the phonon dispersion.[26,27]

X-ray TDS offers the opportunity to answer unresolved questions concerning large-wavevector phonons and about phonons propagating along any direction, as it has sufficiently high photon momentum and can be carried out in experimental geometries which allow the entire Brillouin zone to be probed. Consequently, TDS can illuminate potential anisotropies in the phonon population resulting from the spatial dependence and discretization of phonon modes in the out-of-plane direction.

X-ray TDS has a high sensitivity to the lowest energy acoustic branches of the phonon dispersion because the scattering cross section is proportional to the inverse square of the phonon energy. In the dominant contribution to the TDS signal, termed first-order TDS, the scattered x-ray intensity at each wavevector in reciprocal space is related to the population of phonons, weighted by factors derived from the phonon atomic displacements and x-ray polarization.[29,30] A straightforward geometric construction relates the wavevector of the scattered x-rays in first-order TDS to the phonon wavevector probed: the phonon wavevector is given by the difference between the wavevector of the scattered x-ray and the nearest point on the reciprocal lattice. Synchrotron sources provide high x-ray fluxes, which allow TDS techniques to be extended to extremely small volumes (few $\mu m^3$), compared to orders-of-magnitude larger volumes required for probing the same range of phonon wavevectors by inelastic x-ray scattering studies with energy resolved detectors,[10] and even larger samples required for inelastic neutron scattering.[31]

## II. METHODS

TDS measurements were performed at the Hard X-ray Nanoprobe beamline operated by the Center for Nanoscale Materials at the Advanced Photon Source, Argonne National Laboratory. The experimental arrangement is a further development of a previously described approach,[32,33] which now allows TDS to be probed in a three-dimensional volume of reciprocal space spanning the entire Brillouin zone. X-ray scattering patterns were collected in a transmission geometry in which the incident beam passed through the nanomembrane, at a specific angle with respect to the surface normal. Silicon nanomembranes in the form of freestanding windows were fabricated from thinned silicon-on-insulator,[34,35] resulting in suspended flat membranes with a lateral dimension of 200 μm, supported on a bulk Si frame as illustrated in Fig. 1(a). Further nanomembrane fabrication details are available in the supplemental materials.[36]

The three-dimensional contour plot shown in Fig. 1(b) shows phonon-energy isosurfaces for the transverse acoustic (TA) modes of bulk Si calculated by diagonalizing the dynamical matrix for silicon using bulk elastic constants.[30] The TA branches are the lowest-frequency modes in Si, and provide the dominant contribution to the TDS intensity, due to the consequent high thermal population of phonons. Isosurfaces of different phonon energies are shown in Fig. 1(b), with a section removed in the foreground to reveal the lower energy phonons near the zone center. Fig. 1(b) demonstrates the complexity and symmetry of the bulk Si phonon dispersion. The complex phonon dispersion requires a three-dimensional phonon probe, particularly in nanomaterials where the bulk symmetry does not apply.

A quantitative prediction of the three-dimensional TDS intensity map from a bulk silicon crystal is shown in Fig. 1(c), simulated using the approximation that the scattering is dominated by the first-order TDS process. In Fig. 1(c), a series of surfaces show different levels of predicted TDS intensity throughout the Brillouin zone, with a section removed to reveal the high intensities near the zone center. Fig. 1(c) depicts the predicted TDS intensity in the zone centered on (-3, -1, 1), which is the same zone measured in this experiment. It is important to note that the distribution of the TDS intensity depends on which reciprocal lattice vector the zone is centered on. Fig.1(c) illustrates that a complete analysis of the TDS requires a three-dimensional experimental sampling methodology in which the intensity distribution across multiple slices of reciprocal space is measured. The three-dimensional experimental approach reported here expands beyond previous studies employing a single slice of reciprocal space.[32,33]

The relationship between the wavevectors of phonons and scattered x-rays in first-order TDS produces a geometric condition that is effectively equivalent to the Ewald sphere because the energies of the incident and scattered x-rays are nearly equal. The Ewald sphere construction sweeps through reciprocal space as the incident angle of the x-ray beam with respect to the surface normal direction of the membrane is varied in a series of steps. Repeating the collection of TDS data at each step is equivalent to collecting a series of closely separated slices, each providing a two-dimensional map of TDS intensities, sampling a set of cross-sections spanning the Brillouin zone, as shown in Fig. 1(d). Real-space sample orientations spanning an angular range of 40° are sufficient to probe phonons across the entire Brillouin zone centered on the (-3, -1, 1) reciprocal lattice point. TDS intensity distributions were collected from membranes with thicknesses of 97 and 21 nm and, for comparison, bulk-like several micrometer-thick regions of the silicon frame near the edges of the membranes. The background contribution to the detected intensity arising out of scattering from the x-ray optics was measured by directing the incident beam through a frame in which the membrane had been mechanically removed, and was subtracted from the raw TDS data.

Figure 2 shows TDS intensity distributions for nanomembranes with thicknesses of 97 nm (Fig 2(a)) and 21 nm (Fig 2(b)). The intensity distributions in Figs. 2(a) and (b) show the TDS intensity acquired with different incident angles of the x-ray beam. The two-dimensional sections of reciprocal space corresponding to these images are shown in in Fig. 1(d). The boundaries of the Brillouin zones are plotted as solid orange lines superimposed on the intensity data in Fig. 2. A three-dimensional representation of the TDS intensity distribution constructed from the detector images is also shown for each thickness. The recorded intensities at each angular setting are multiplied by an angle-dependent factor to account for the difference in volume of the sample probed in each geometry. Intense scattering is observed (i) near the zone center, due to the large population of small-wavevector phonons, and (ii) along diffuse streaks close to high-symmetry directions where phonon modes have low frequencies. The three-dimensional representation extracted from the detector images qualitatively reproduces the predicted shape of the TDS intensity distribution shown in Fig. 1(c).

### III. RESULTS

The well-defined relationship between the wavevector of scattered x-rays and crystal momentum within the Si membrane allows us to directly interpret the intensity distribution as a function of the phonon wavevector. The TDS intensity along arbitrary crystallographic directions was extracted by numerically resampling the intensity onto a three-dimensional grid spanning the Brillouin zone and plotted as a function of the reciprocal space vector connecting each point to the (-3, -1, 1) zone center. The experimental and analytical techniques presented here represent an advance over previous TDS measurements because they allow the extraction of intensity profiles along arbitrary crystallographic directions, including the commonly described high-symmetry directions. Intensity distributions along <100>, <110>, and <111> high-symmetry directions, are shown in Fig. 3 for the 97 nm and 21 nm membranes. The [010], [-1,1,1], and [1,1,-1] directions are not shown in Fig. 3 because there is a lower density of points sampling the intensity along those directions as a result of the small number of Ewald sphere slices intersecting those paths. The intensity axis in Figure 3 is normalized by the membrane thickness to allow the profiles for the different crystallographic directions to be compared. The predicted TDS intensity based on the bulk Si phonon dispersion is scaled by a single parameter to fit the experimental data along several slices of reciprocal space in Fig. 3. These fits show that the distribution of the TDS intensity expected from bulk Si matches the observed experimental data, and also indicate that the TDS from the thin membrane is more intense than expected based on its thickness.

The variation of TDS intensity with membrane thickness can be explored precisely by plotting the TDS intensity along directions separated by fine angular steps, shown in Figs. 4(a) and (b). Figure 4 shows the ratio of the TDS intensity acquired from the 21 nm membrane to the intensity from the bulk-like 97 nm membrane along these directions. Previous studies have shown that the intensity of scattering from membranes at approximately 100-nm thicknesses follows the scaling proportional to thickness expected for bulk dispersions.[32,33] If the intensity of scattering from the 21 nm membrane was proportional to the thickness of the nanomembrane, the intensity ratios in Fig. 4(a) would exhibit no dependence on direction and have a constant value of 0.22, as indicated by the horizontal dashed lines in Figs 4(a) and (b). Instead, the dependence of the TDS intensity on reciprocal space direction is dramatically different for the two nanomembrane samples, as seen from the anisotropy in the intensity ratio, especially for wavevectors shorter than half the zone radius (or wavelengths larger than a few nanometers).

Figure 4 shows that the phonon populations at wavevectors along in-plane and out-of-plane directions, which would be equal in bulk samples, are different for the smaller nanomembrane thickness. As $Q_z$ increases, the ratio of the observed scattered intensity to the value expected by scaling the intensity from the 97 nm membrane increases. At its maximum, the ratio of the TDS intensities for the 21 nm and 97 nm thicknesses a factor of 4 larger than the value expected from the linear scaling observed in thicker samples and a factor of 2 larger than the in-plane ratio. This result indicates that there is an anisotropy in the TDS intensity from the 21 nm membrane that does not occur in the bulk or in the 97 nm membrane thickness. The in-plane series of directions, Fig. 4(b), does not exhibit strong a dependence on the profile direction, and the intensity distribution in the in-plane directions is the same in both 21 nm and 97 nm membranes. TDS ratio plots from the two bulk-like membrane frames (see supplemental material) show that bulk Si does not exhibit the directional anisotropy observed in nanomembranes.[36]

It is also apparent in Figs. 3 and 4 that the 21 nm membrane produces a higher TDS intensity per unit thickness than the 97 nm membrane across the entire Brillouin zone. There are several possible causes for increased intensity including contributions from a relative increase in the static background scattering or a relative increase in the importance of scattering from the membrane surface, which would contribute a larger fraction of the total scattering observed from the 21 nm-thick membrane sample.

Existing experimental and theoretical results provide some insight into modes with in-plane wavevectors. Brillouin light scattering measurements, while sensitive to out-of-plane atomic displacements, are limited to the measurement of in-plane phonon wavevectors.[5,20,21,23,24] These measurements show evidence of the conversion of bulk transverse and longitudinal acoustic modes into flexural and dilatational modes as predicted by elastic continuum theory, which is accompanied by the development of a quadratic dispersion near the zone center and higher order quantized modes.[20,24,26–28,37,38] The in-plane dispersions of the flexural and dilatational modes of the lowest energy out-of-plane modes have lower energies than the bulk transverse and longitudinal modes in Si.[32,33] The low-energy vibrational modes of thin Si sheets change the total phonon density of states and thus modify total phonon population probed in TDS experiments.[26,27] The introduction of lower-energy phonon modes would contribute to the overall rise in TDS intensity observed in the scattering from the 21 nm-thick membrane sample through an increase in low-frequency phonon population with a correspondingly high x-ray scattering cross-section.[32,33]

There is far less insight available into the out-of-plane phonon mode energies, for which $Q_x=0$. Elastic calculations predict a quantization of modes in the out-of-plane direction due to the finite thickness in that direction. The quantized modes in the out-of-plane direction have wavevectors given by $Q_z=n\pi/d$, where $d$ is thickness of the membrane. Accordingly, the out-of-plane phonon modes can no longer be thought of as having a continuous dispersion.[26] The in-plane wavevector remains continuous, as in the bulk dispersion.[26,27] However the dispersion of out-of-plane acoustic phonons predicted by elastic continuum theory is linear with a slope dependent on the elastic constants, which take on their bulk values in elastic calculations.[26,27] These elastic theory predictions do not predict the development of new out-of-plane modes.

A possible explanation for the higher phonon population in out-of-plane modes is a softening of the elastic constants of the membrane in comparison to the bulk. Softening would lead to a decrease in the frequencies of out-of-plane vibrational modes, with a larger reduction of the elastic constants in the out-of-plane direction than in the in-plane direction. Mechanical studies of ultrathin silicon suggest that the elastic properties are also affected by the reduced dimensionality of the membrane.[39,40] Experimental studies of nanoscale silicon beams report a decrease in the effective Young's Modulus in thin Si sheets. For example, Bhaskar *et al.* report a decrease from 160 GPa to 108 GPa when the thickness was reduced from 200 nm to 50 nm.[41] The reduction in the effective elastic modulus was attributed to surface effects, the presence of the native oxide, and fabrication-induced defects.[42] This effect is not captured in the elastic continuum calculations discussed above because elastic predictions rely on the bulk elastic constants.

We note, however, that continuum theories have limited validity in the large-wavevector regime and that a full comparison will require a more complete atomistic picture. A limited number of molecular dynamics calculations have been performed for membranes as thick as 20 nm.[5,6] The phonon dispersion calculated by molecular dynamics extends into the large-wavevector regime and are consistent with elastic continuum theory, but has only been presented for the in-plane directions.[5,6] Molecular dynamics calculations show evidence of an average softening of the elastic moduli, especially for modes that create out-of-plane atomic displacements, which would be consistent with anisotropy in the softening of the elastic constants.[6]

## IV. CONCLUSIONS

Synchrotron x-ray thermal diffuse scattering is used to generate three-dimensional maps of phonon population throughout the entire Brillouin zone in nanoscale silicon. Nanomembranes with 21 nm thickness show an increased population in modes with out-of-plane phonon wavevectors, even for phonon wavelengths as small as 2 nm. A comparison of the scattered intensities from different thicknesses of silicon nanomembranes shows that the intensity deviates from the expected thickness scaling and that an anisotropy in the TDS intensity is observed at a thickness of 21 nm that possesses the same symmetry as the sample geometry. These deviations from bulk results could be explained by the emergence of new lower energy modes across the entire Brillouin zone and a modification of the elastic constants resulting in a higher phonon population. We further note that more detailed comparisons will require extending the existing theory of TDS to the nanoscale to account for the differences in atomic displacements in confined modes. Thermal diffuse scattering approaches provide means to study the phonon populations across the entire Brillouin zone of nanoscale crystalline materials. These results will directly enable 3-dimensional visualization of phonon band engineering in such advanced nanoscale structures as those fabricated for phononic, thermoelectric and acousto-optic applications.[3,7,8]

# ACKNOWLEDGEMENTS

GG, KM, and PE acknowledge support from the US Air Force Office of Scientific Research, through contract FA9550-10-1-0249 for the development of diffuse scattering methods and from the U.S. DOE, Basic Energy Sciences, Materials Sciences and Engineering, under contract no. DE-FG02-04ER46147 for the development of three-dimensional x-ray scattering analysis techniques. Nanomembranes were fabricated at the Wisconsin Center for Applied Microelectronics at the University of Wisconsin-Madison, supported in part by the UW MRSEC (NSF DMR-1121288), and at the Center for Nanoscale Materials at Argonne National Laboratory. The Center for Nanoscale Materials, an Office of Science user facility, is supported by the U. S. Department of Energy, Office of Science, Office of Basic Energy Sciences, under Contract No. DE-AC02-06CH11357. Use of the Advanced Photon Source was supported by the U. S. Department of Energy, Office of Science, Office of Basic Energy Sciences, under Contract No. DE-AC02-06CH11357.

McElhinny *et al.* FIG. 1.

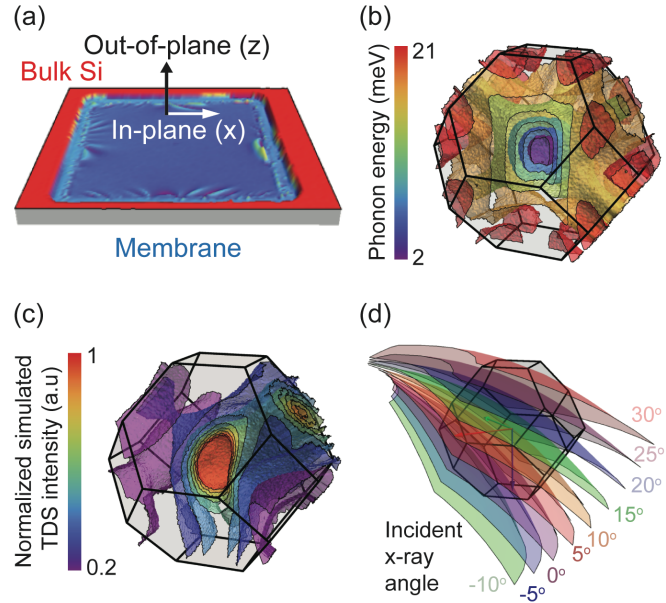

**FIG. 1. Silicon nanomembrane phonon dispersion and x-ray thermal diffuse scattering.** (a) Suspended 21-nm-thick silicon membrane. The thin dimension of the membrane is along the out-of-plane (*z*) direction. (b) Bulk phonon dispersion of Si, with a sector of the plot removed to expose low-phonon-energy contours. (c) Simulated distribution of x-ray thermal diffuse scattering (TDS) intensity inside the (-3 -1 1) zone using the bulk Si phonon dispersion. A sector of the prediction has been removed to expose regions of high intensity. High intensity streaks occur along <111> and <001> directions, as is apparent in the phonon dispersion. (d) X-ray TDS collects information about the phonon population along the two-dimensional surface of the Ewald sphere, which is swept through the Brillouin zone by rotating the sample through a series of values of the incident angle of the x-ray beam.

McElhinny *et al.* FIG. 2.

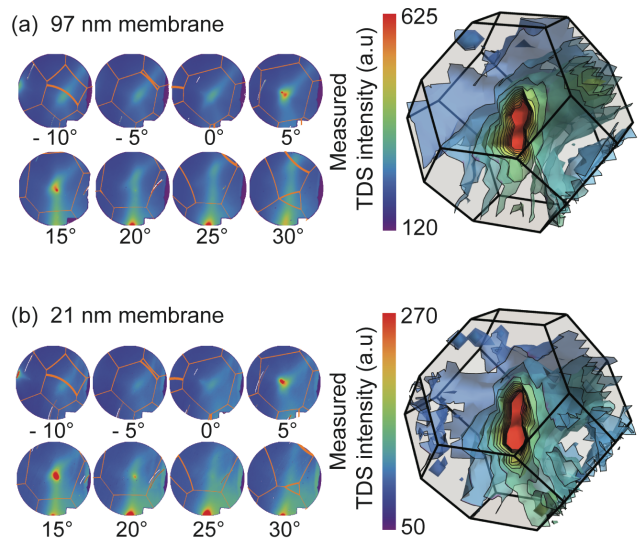

**FIG. 2. TDS intensity distributions for silicon nanomembranes.** Scattering patterns with thicknesses of (a) 97 nm (b) 21 nm, acquired with incident angles from -10° to 30° in 5° steps. The image acquired at an incident angle of 10° excites the highly intense (-3, -1, 1) Bragg reflection of the nanomembrane, saturating the detector, and is thus not shown. The crystal truncation rod of the Si sheet appears as an intense highly localized feature in several images. The three-dimensional representation of the experimental TDS intensity constructed from the detector images appears in the right panels of (a) and (b). The data has been removed from a sector of the Brillouin zone in order to show the high-intensity contours near the zone center.

McElhinny *et al.* FIG. 3.

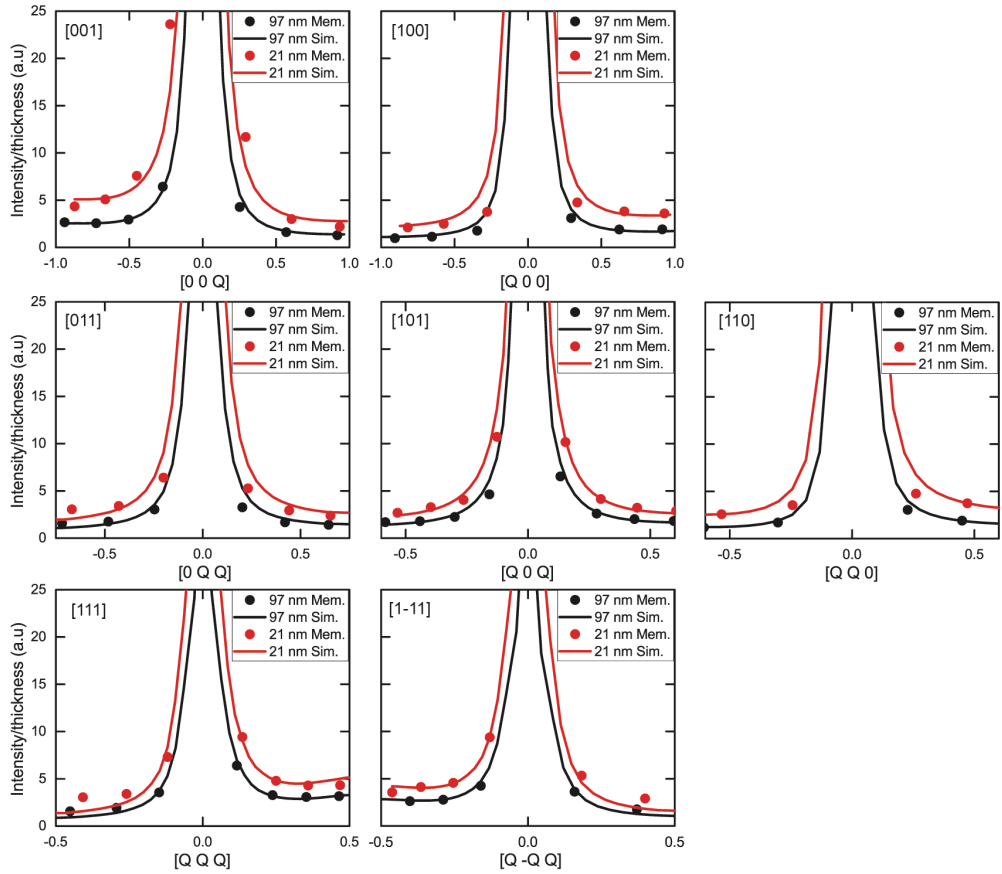

**FIG. 3. High-symmetry-direction profiles of the TDS intensity.** The intensity is normalized by x-ray data acquisition time and membrane thickness. Black and red dots represent measured intensities from nanomembranes with thicknesses of 97 nm and 21 nm, respectively. Black and red lines are the first-order TDS intensity simulated using the bulk phonon dispersion in which the overall intensity is multiplied by a single overall scaling parameter to fit the corresponding experimental intensities.

McElhinny *et al.* FIG. 4.

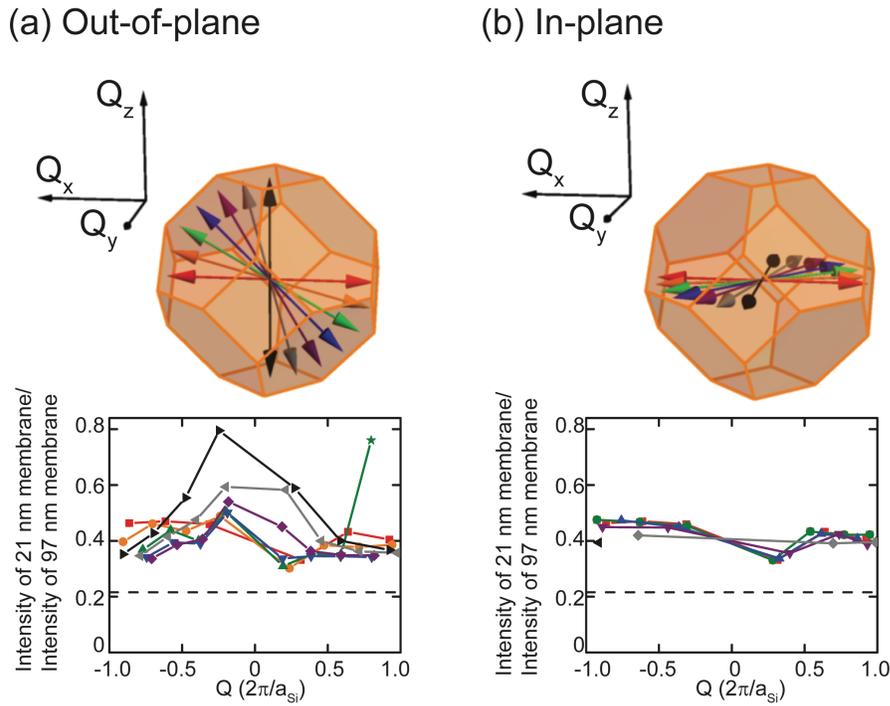

**FIG. 4. Anisotropy of nanomembrane thermal diffuse scattering.** Ratios of intensities from 21 nm-thick and 97 nm-thick silicon nanomembranes along (a) out-of-plane and (b) in-plane directions. The directions of the profiles are shown in the corresponding color in the inset schematic of the Brillouin zone. The anomalously intense point plotted with a star includes an artifact due to powder diffraction from the x-ray optics and thus does not fit the overall trend. The horizontal dashed line indicates the expected ratio of the intensities of TDS from the 21 nm-thick and 97 nm-thick membranes based on the linear scaling of the intensity with thickness.

# Three-dimensional phonon population anisotropy in silicon nanomembranes

Kyle M. McElhinny[1], Gokul Gopalakrishnan[1†], Martin V. Holt[2], David A. Czaplewski[2], Paul G. Evans[1*]

[1] Materials Science and Engineering, University of Wisconsin-Madison, Madison, WI 53706, USA

[2] Center for Nanoscale Materials, Argonne National Laboratory, Argonne, IL 60439, USA

* Corresponding author. E-mail: pgevans@wisc.edu

†Present address: Engineering Physics, University of Wisconsin-Platteville, Platteville, WI 53818, USA

**Supplementary Materials:**

**Materials and Methods**

A 10 keV x-ray beam was focused using a capillary condenser to a spot with a diameter of 30 μm. The scattered x-ray data was collected using a charge-coupled device (Mar-165, MAR, Inc.) with a pixel size of 82 μm, corresponding to a wavevector resolution of $1 \times 10^{-3}$ Å$^{-1}$ in scattering patterns. Nanomembrane fabrication followed the using the edge-induced flattening method described in ref. [34]. The dip in the membrane profile as it leaves the ledge as shown schematically in Fig. 1(a) may be due to an artifact of optical interferometry [40].

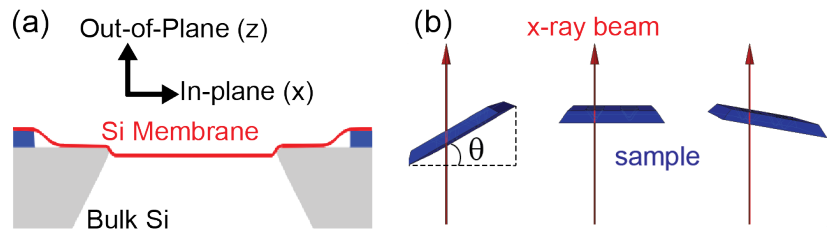

**FIG. S1. X-ray scattering methods.** (a) Cross section of a suspended silicon membrane. The thin dimension of the membrane is termed the out-of-plane *z* direction. (b) Scattering geometry

indicating the rotation of the sample through the variation of the incident angle $\theta$ of the x-ray beam.

**Absence of Anisotropy in Bulk Si TDS**

Figure S2 shows the ratio of intensity between the bulk frame of the 21 nm membrane and the frame of the 97 nm membrane. The in-plane and out-of-plane directions are plotted as was done for the membranes in figure 4. Neither the in-plane or out-of-plane direction profiles show evidence of the anisotropy observed in the membrane. The in-plane and out-of-plane ratios are slightly below one due to artifacts arising from differences in the total thickness sampled for the bulk measurement and from small differences in the sample orientation that lead to slightly different regions being sampled in reciprocal space with slightly different phonon populations. The feature at $Q=0$ in the bulk ratios in Figure 4(c) results from differences in the sample geometry, leading to a small error in the sample wavevector being compared.

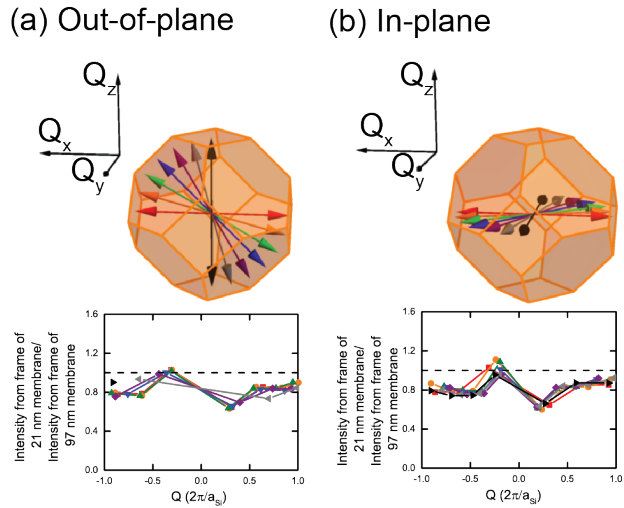

**FIG. S2. Absence of anisotropy in bulk Si.** The ratio of intensity between the bulk frame of the 21 nm membrane and the frame of the 97 nm membrane. The (a) out-of-plane and (b) in-plane

directions are plotted as was done for the membranes in figure 4. Neither the in-plane or out-of-plane direction profiles show evidence of the anisotropy observed in the membrane.

**Supplemental References**